\documentclass[aps,showpacs,twocolumn]{revtex4}
\usepackage{amsmath}
\usepackage{amssymb}
\usepackage{amscd}
\usepackage{wasysym}
\usepackage[ansinew]{inputenc}
\usepackage[T1]{fontenc}
\usepackage{ae,aecompl}
\usepackage[dvips]{graphicx}
\usepackage{color}
\newcommand{\ri}{{ \rm i }}
\newcommand{\re}{{ \rm e }}
\newcommand{\rd}{{ \rm d }}

\newcommand{\be}{\begin{equation}}
\newcommand{\ee}{\end{equation}}

\begin{document}
\bibliographystyle{apsrev}
\title{Semiclassical quantization of an $N$-particle Bose-Hubbard model}
\author{E. M. Graefe and H. J. Korsch}
\email{korsch@physik.uni-kl.de}
\affiliation{FB Physik, Technische Universit{\"a}t Kaiserslautern,
D-67653 Kaiserslautern, Germany}
\date{\today }

\begin{abstract}
A semiclassical Bohr-Sommerfeld approximation is derived for an
$N$-particle, two-mode Bose-Hubbard system modeling a Bose-Einstein
condensate in a double-well potential. This semiclassical
description is  based on the `classical' dynamics
of the mean-field Gross-Pitaevskii equation and is expected to be valid for 
large $N$. We demonstrate the possibility to reconstruct quantum properties 
of the $N$-particle system from the mean-field dynamics. The
resulting semiclassical eigenvalues and eigenstates are found to be
in very good agreement with the exact ones, even for small values of $N$,
both for subcritical and supercritical particle interaction strength where tunneling has to be
taken into account.
\end{abstract}

\maketitle

\section{Introduction}
Even for weakly interacting particles,
a full many-body treatment of Bose-Einstein condensates (BEC)
is only possible for a small number $N$ of particles. Most often
a mean-field approximation is used, which describes
the system quite well for large $N$ at low temperatures. In this mean-field 
approach, the bosonic field operators are replaced by c-numbers, the condensate
wavefunctions. This constitutes a classicalization and therefore
the result of the mean-field approximation, the Gross-Pitaevskii equation 
(GPE), is often denoted as `classical', despite of the fact that the
GPE is manifestly quantum, i.e.~it reduces to
the usual linear Schr\"odinger equation for vanishing
interparticle interaction. 
Therefore, in order to avoid misunderstanding, this 
inversion of the second quantization may be called a second classicalization.

In a number of recent papers, consequences of the classical nature of the mean-field 
approximation are discussed and semiclassical aspects are introduced.
For a two-mode Bose-Hubbard model,
Anglin and Vardi \cite{Vard01b,Angl01} consider
equations of motion which go beyond the standard mean-field theory
by including higher terms in the Heisenberg equations of motion.
The classical-quantum correspondence has been studied in terms of phase space
(Husimi) distributions \cite{Mahm05} for such systems. 
Mossmann and Jung \cite{Moss06} demonstrate for a
three-mode Bose-Hubbard model that
the organization of the $N$-particle eigenstates
follows the underlying classical, i.e.~mean-field, dynamics.
A generalized Landau-Zener formula for the mean-field description of interacting
BECs in a two-mode system has been derived by studying the many particle system \cite{06zener_bec}.
In \cite{Wu06} the commutability between the classical and the adiabatic limit for the same system is studied and first steps towards a semiclassical treatment of the problem are reported.

The purpose of the present paper is to show that the
mean-field model is not only capable to approximate the
interacting $N$-particle system in the limit of large $N$
and to allow for an interpretation of the organization of
the $N$-particle eigenvalues and eigenstates, but can also be
used to reconstruct approximately the individual eigenvalues and 
eigenstates in a semiclassical Bohr-Sommerfeld (or EBK) manner with astounding 
accuracy even for a small number of particles.
This will be demonstrated for
$N$  bosonic particles in a two-mode system,
a many-particle Bose-Hubbard Hamiltonian,
describing for example
the low-energy dynamics of a BEC in a (possibly asymmetric) double-well potential.
This system is related to a classical non-rigid pendulum in the mean-field approximation
(see, e.g., \cite{Mahm05} and references therein).

Both the many particle model and its
classical version -- which is often denoted as the discrete self-trapping equation 
for reasons which will become obvious in the following --
have been studied for decades in different context (see \cite{semiMP_Bem2}). 
A detailed semiclassical analysis is, however, missing up to now. Previous
work in this direction concentrated on the symmetric case, 
where the permutational symmetry of the system with respect to an interchange of
the two modes simplifies an analysis. Semiclassical expressions
for the tunneling splittings of the eigenvalues have been derived 
\cite{Enz86,Scha87} in context of the spin-system in eqn.~(\ref{BH-hamiltonian-SR}) below
(see also \cite{Bern90,Gara91} for a perturbative treatment of the splittings and
\cite{Fran01} for a detailed analysis of the quantum spectrum).

In the following we will first give a short overview of the many particle description
of the model, the celebrated mean-field approximation and their correspondence. 
Afterwards we focus on the question to which extent
many particle properties can be extracted from the mean-field
system by an inversion of this `classical' approximation in a
semiclassical way using the EBK-quantization method \cite{Brac97}.

\section{Two-mode Bose-Hubbard model and mean-field approximation}
In a two-mode approximation at low temperatures a BEC in a double-well
potential can be described by a second quantized many particle Hamiltonian of Bose-Hubbard type:
\be
 \hat H= \varepsilon (\hat n_1- \hat n_2)
 + v(\hat a_1^\dagger \hat a_2 + \hat a_2^\dagger \hat a_1)
 +g\left(\hat n_1^2+\hat n_2^2 \right)\,.
 \label{BH-hamiltonian}
\ee
Here $\hat a_j$, $\hat a_j^\dagger$ are bosonic
particle annihilation and creation operators for the jth mode and
$\hat n_j = \hat a_j^\dagger\hat a_j$ are the mode number operators.
The mode energies
are $\pm\varepsilon$, $v$ is the coupling constant and $g$ is the
strength of the onsite interaction.
In order to simplify the discussion, we assume here that $v$ is positive 
and $g$ is negative \cite{semiMP_Bem1}.
The Hamiltonian (\ref{BH-hamiltonian}) commutes with the total number operator
$\hat N=\hat n_1 + \hat n_2$ 
and the number $N$ of particles, the eigenvalue of $\hat N$,
is conserved. For a given value of $N$, we have $N+1$
eigenvalues of the Hamiltonian (\ref{BH-hamiltonian}).
Alternatively, the system can be described in the Schwinger representation
by transformation to angular momentum operators
$\hat J_x=( \hat a_1^\dagger \hat a_2+\hat a_2^\dagger \hat a_1)/2$\,,
$\hat J_y=( \hat a_1^\dagger \hat a_2-\hat a_2^\dagger \hat a_1)/2\ri$\,,
$\hat J_z=( \hat a_1^\dagger \hat a_1-\hat a_2^\dagger \hat a_2)/2$\,.
The Hamiltonian (\ref{BH-hamiltonian}) then takes the form
\be
\hat H=2\varepsilon\hat J_z+2v\hat J_x+2g(\hat J_z^2+N^2/4),
\label{BH-hamiltonian-SR}
\ee
where the total angular momentum is $J=N/2$.

\begin{figure}[t]
\centering
\includegraphics[width=6.6cm]{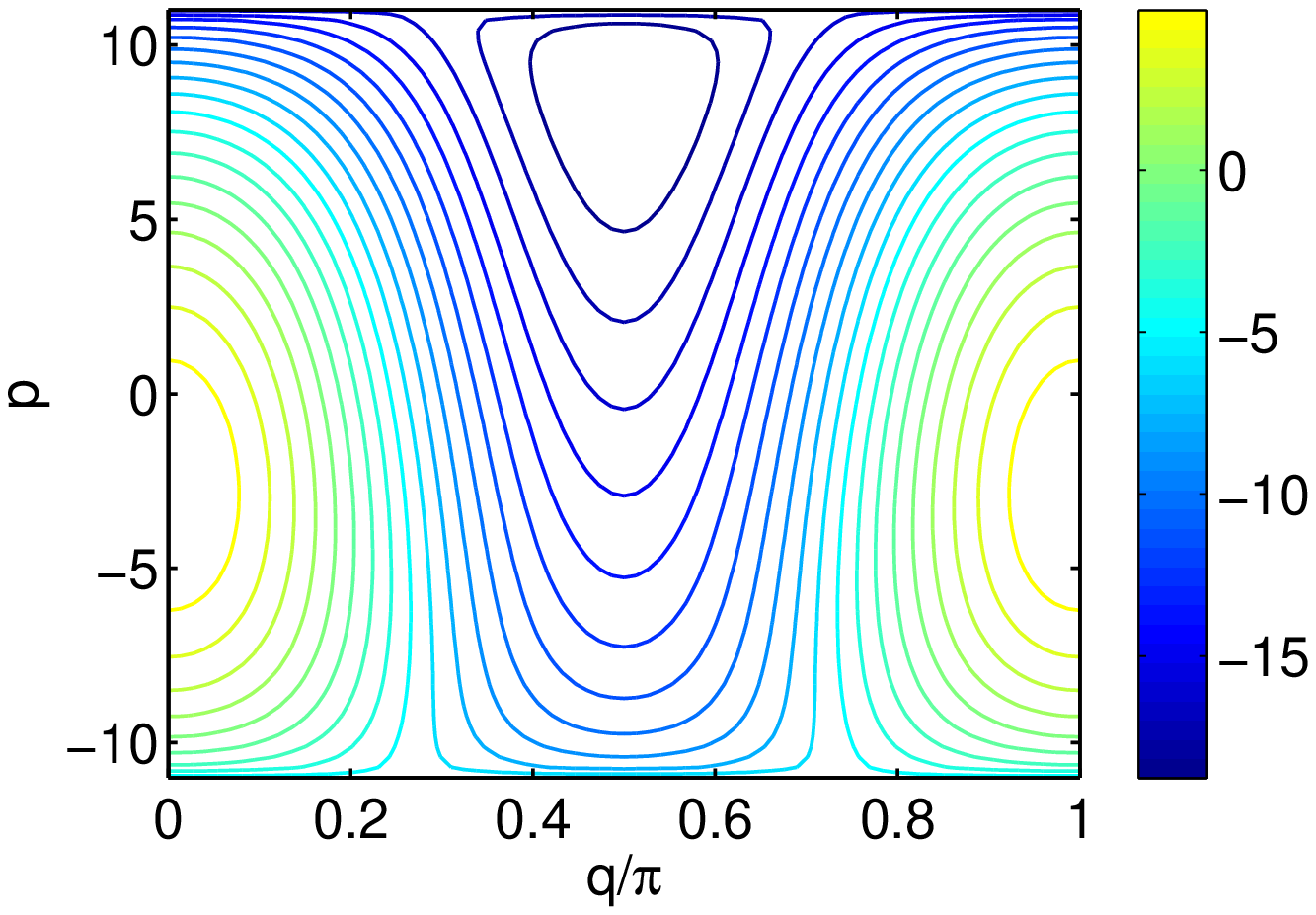}
\includegraphics[width=6.6cm]{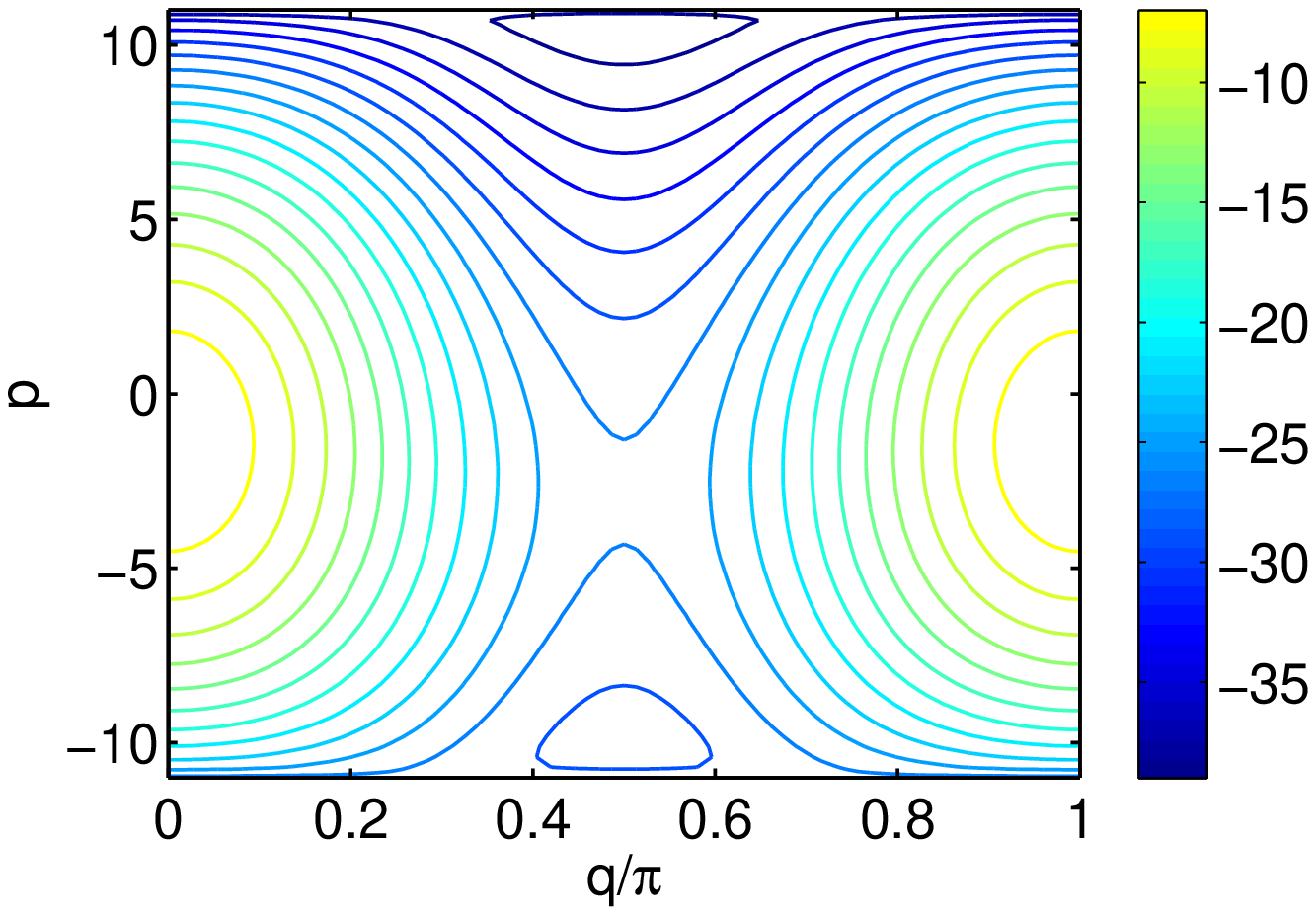}
\caption{\label{fig-phasespace}
(Color online) Phase space portrait of the mean-field Hamiltonian ${\cal H}(p,q)$ in (\ref{klHam})
for $v=1$ and $\varepsilon=-0.5$ in the subcritical ($g=-1/N_s$, top) and supercritical
($g=-3/N_s$, bottom) region for $N=10$ and $\hbar=1$.}
\end{figure}

The celebrated mean-field description can be most easily
formulated as a replacement of operators by c-numbers
$\hat a_j\rightarrow \psi_j$, $\hat a_j^\dagger \rightarrow \psi_j^*$.
Since the c-numbers commute in contrast
to the quantum mechanical operators, the
transition quantum $\to$ classic and vice versa is not one-to-one. To avoid ambiguities
one has to replace symmetrized products of the operators
by the corresponding products of c-numbers.
Therefore we will start on the many particle side with
a symmetrized Bose-Hubbard Hamiltonian in the following, where the
$\hat n_j$
are replaced by $ \hat n^s_j=(\hat a_j^\dagger\hat a_j+\hat a_j\hat a_j^\dagger)/2$
(see also \cite{Moss06}). This symmetrization affects only
the nonlinear term in (\ref{BH-hamiltonian}) and the symmetrized $\hat H$ is related to (\ref{BH-hamiltonian}) by an
additive constant term depending only on $\hat N$. 
Note that thus the number operator $\hat N=\hat n_1 +\hat n_2= \hat n^s_1+ \hat n^s_2 -\hat 1$ 
is replaced by $\vert \psi_1\vert^2+\vert\psi_2\vert^2-1$ 
and therefore the mean-field wavefunction is normalized as 
$\vert \psi_1\vert^2+\vert\psi_2\vert^2=N+1=N_s$.

The mean-field time evolution is given by the two level nonlinear Schr\"odinger equation, resp.~GPE,
\be\label{2stGPE}
  \ri \hbar \,\frac{\rd}{\rd t}
  \begin{pmatrix} \psi_{1} \\ \psi_2 \end{pmatrix}=
\begin{pmatrix}
  \varepsilon + 2g |\psi_1|^2 & v \\ v & -\varepsilon + 2g |\psi_2|^2
   \end{pmatrix}
  \begin{pmatrix} \psi_{1} \\ \psi_2 \end{pmatrix}\,,
\ee
where $\psi_1$ and $\psi_2$ are the amplitudes of the two condensate modes.

The Schr\"odinger equation, linear or nonlinear, 
has the canonical structure of classical dynamics \cite{Ablo04,Fadd87,Hesl85}:
The time dependence of the complex valued
mean-field amplitudes can be written as
canonical equations of motion
\be\label{can-eqn-psi}
\ri\hbar\frac{\rd \psi_j}{\rd t}=\frac{\partial {\cal H}}{\partial \psi_j^*}\quad \text{and}\quad \ri\hbar\frac{\rd \psi_j^*}{\rd t}=-\frac{\partial {\cal H}}{\partial \psi_j}
\ee 
with a Hamiltonian function ${\cal H}=\epsilon(|\psi_1|^2-|\psi_2|^2)+v(\psi_1^*\psi_2+\psi_1\psi_2^*)+g(|\psi_1|^4+|\psi_2|^4)$. 
The conservation of the particle number 
allows a reduction of the dynamics
to an effectively one-dimensional Hamiltonian evolution
by an amplitude-phase decomposition
$\psi_j=\sqrt{n_j+1/2}\,\re^{\ri q_j}$ in terms of the
canonical coordinate $q=(q_1-q_2)/2$, an angle, and the
(angular)momentum $p=(n_1-n_2)\hbar$, with the Hamiltonian function
\be\label{klHam}
{\cal H}(p,q) = \varepsilon \,\frac{p}{\hbar}+v\sqrt{N_s^2-\frac{p^2}{\hbar^2}}\,\cos(2q)
+\frac{g}{2}\big(N_s^2+\frac{p^2}{\hbar^2}\big)\,,
\ee
where $N_s$ is the normalization of $\psi$.
Introducing the new variables the canonical equations of motion~(\ref{can-eqn-psi}) 
obtain their usual appearance  $\dot{q}=\partial {\cal H}/\partial p$ and $\dot{p}=-\partial {\cal H}/\partial q$:
\begin{align}
\dot{p}=&2v\sqrt{N_s^2-\frac{p^2}{\hbar^2}}\sin(2q)\\
\dot{q}=&\frac{\varepsilon}{\hbar}-v\frac{p}{\hbar^2\sqrt{N_s^2-\frac{p^2}{\hbar^2}}}\,\cos(2q)
+g \frac{p}{\hbar^2}.
\end{align}
This describes the classical dynamics of a non-rigid pendulum
where the phase space is finite, $-N_s\hbar\le p\le N_s\hbar$, $0\le q\le \pi$,
if the lines $q=0$ and $q=\pi$ are identified.

One of the prominent features of the two-mode system
is the self-trapping effect, which leads to the emergence of additional stationary states favoring one of the wells above a critical particle interaction strength. For a discussion of the relation between mean-field
and $N$-particle behavior see, e.g., \cite{Aubr96,Milb97} and references therein
and \cite{Holt01a} for its control by external driving fields.
The self-trapping transition occurs at $g=-v/N_s$ and
is connected to a bifurcation
of the stationary states, the fixed points
of the  Hamiltonian (\ref{klHam}), in the mean-field approximation. 
Figure \ref{fig-phasespace} shows phase space portraits of ${\cal H}(p,q)$
for sub- and supercritical particle interaction strength.
In the subcritical region one has
a maximum, $E^+$, at $q=0$ and a minimum, $E^-$,
at $q=\pi/2$. For the symmetric case  $\varepsilon=0$, both are located at $p=0$, which means that 
the population in both wells is the same. In the supercritical region
the minimum bifurcates into two
minima, $E^-_\pm$, and a saddle point,
$E^-_{\rm saddle}>E^-_\pm$. Even for the case $\varepsilon=0$ the two minima are located at a finite value of
the population imbalance. In these stationary states the condensate mainly populates one of the wells, which leads to the name {\it self-trapping}.
In phase space, the regions with oscillations around one of the two
minima are separated by a separatrix passing through the saddle point.
The period of the separatrix motion is infinite.

\begin{figure}[b]
\centering
\includegraphics[width=6.6cm]{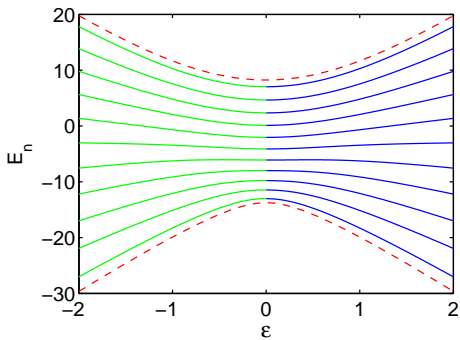}
\caption{\label{fig-levels_mp_mf-1}
(Color online) Many particle energies $E_n$ and mean-field eigenenergies ${\cal H}_{stat}$ (\textcolor{red}{- -})
as a function of the onsite energy $\varepsilon$
in the subcritical  region ($g=-0.5/N_s$) for $v=1$ and $N=10$ particles ($\hbar=1$).
The spectrum is organized by the mean-field eigenenergies ${\cal H}_{stat}$ (\textcolor{red}{- -}).
The exact quantum eigenvalues shown for $\varepsilon\leq 0$ (\textcolor{green}{---}) are almost exactly reproduced by the
semiclassical ones, $E_n^{sc}$, shown for $\varepsilon\geq 0$ (\textcolor{blue}{---}).}
\end{figure}

\begin{figure}[t]
\centering
\includegraphics[width=4.2cm]{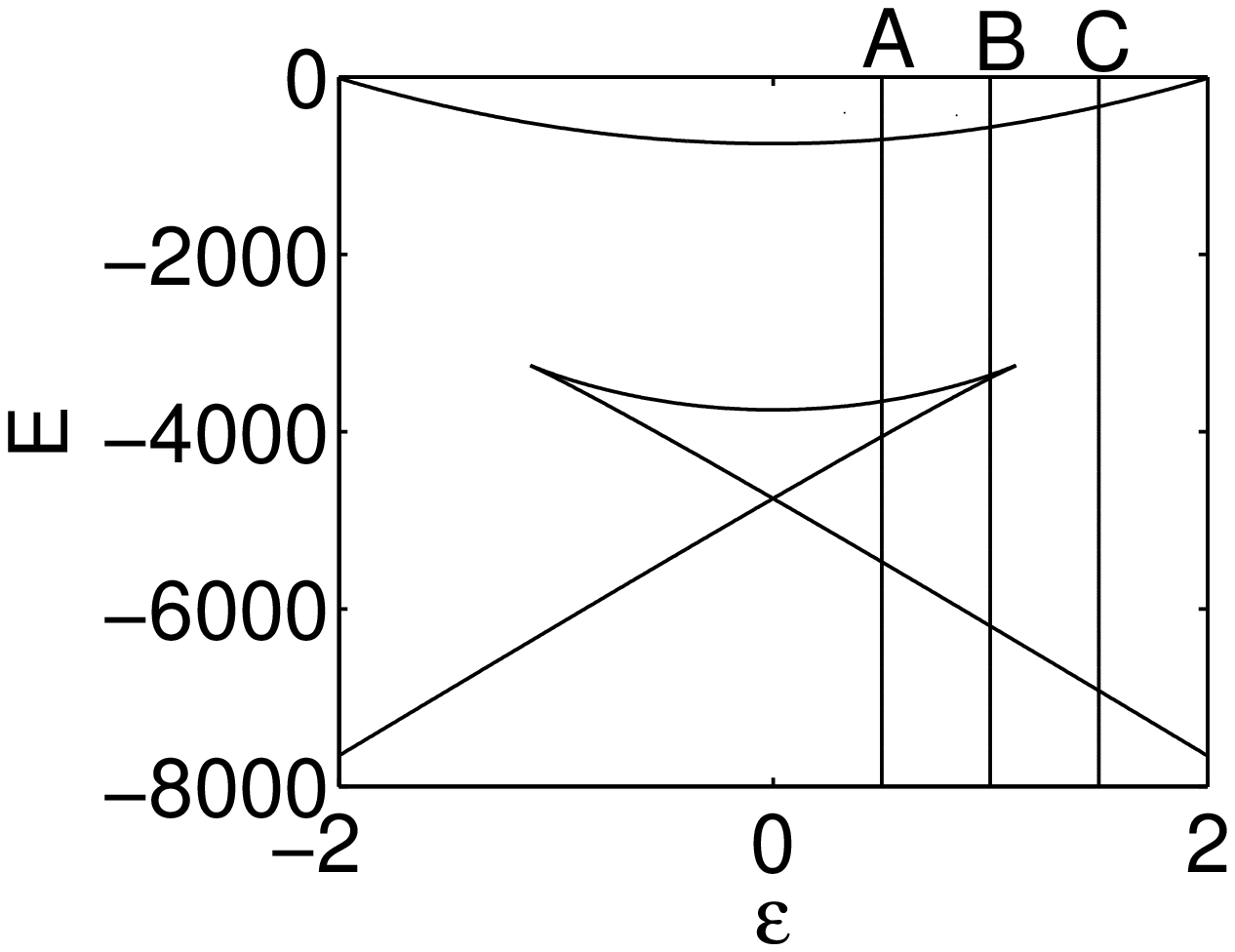}
\includegraphics[width=4.2cm]{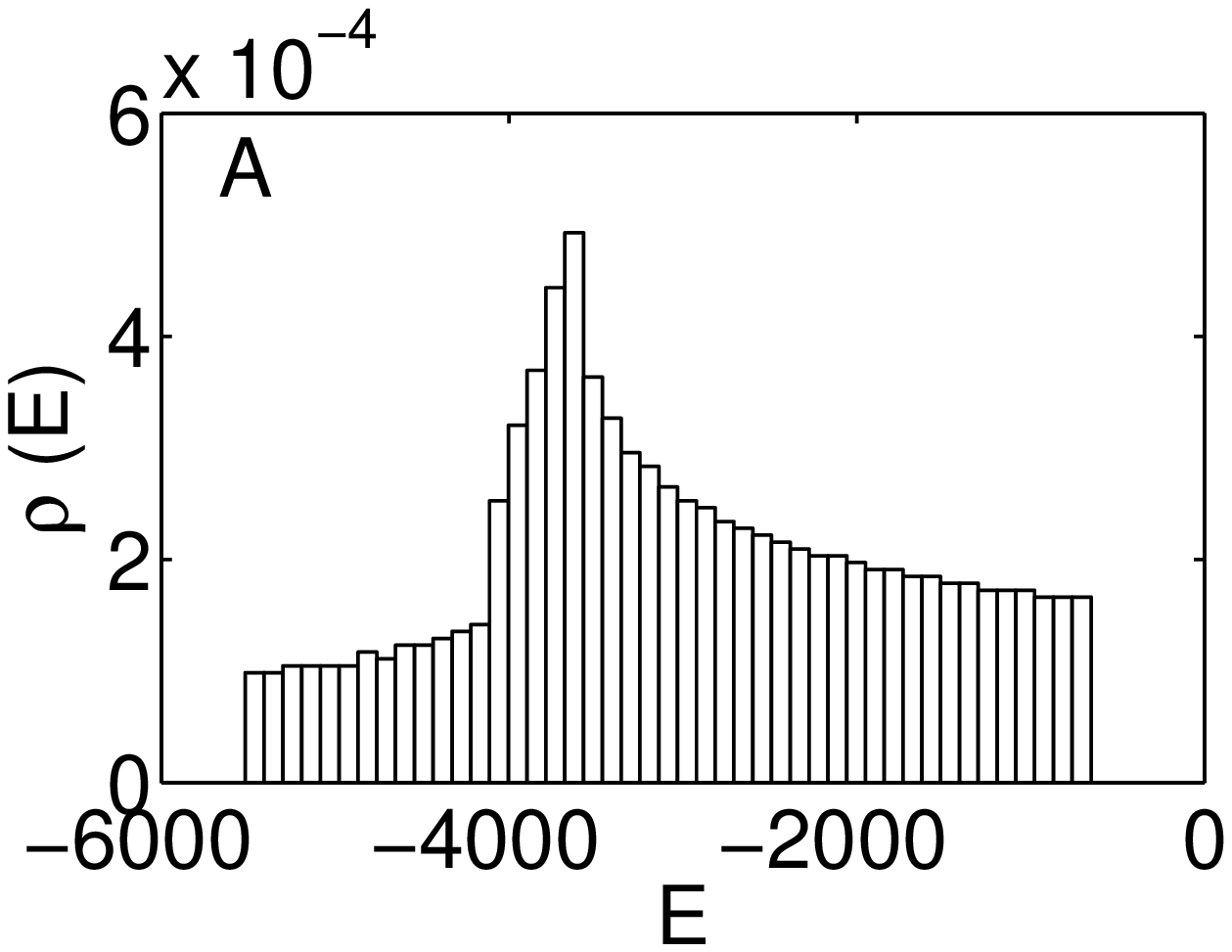}
\includegraphics[width=4.2cm]{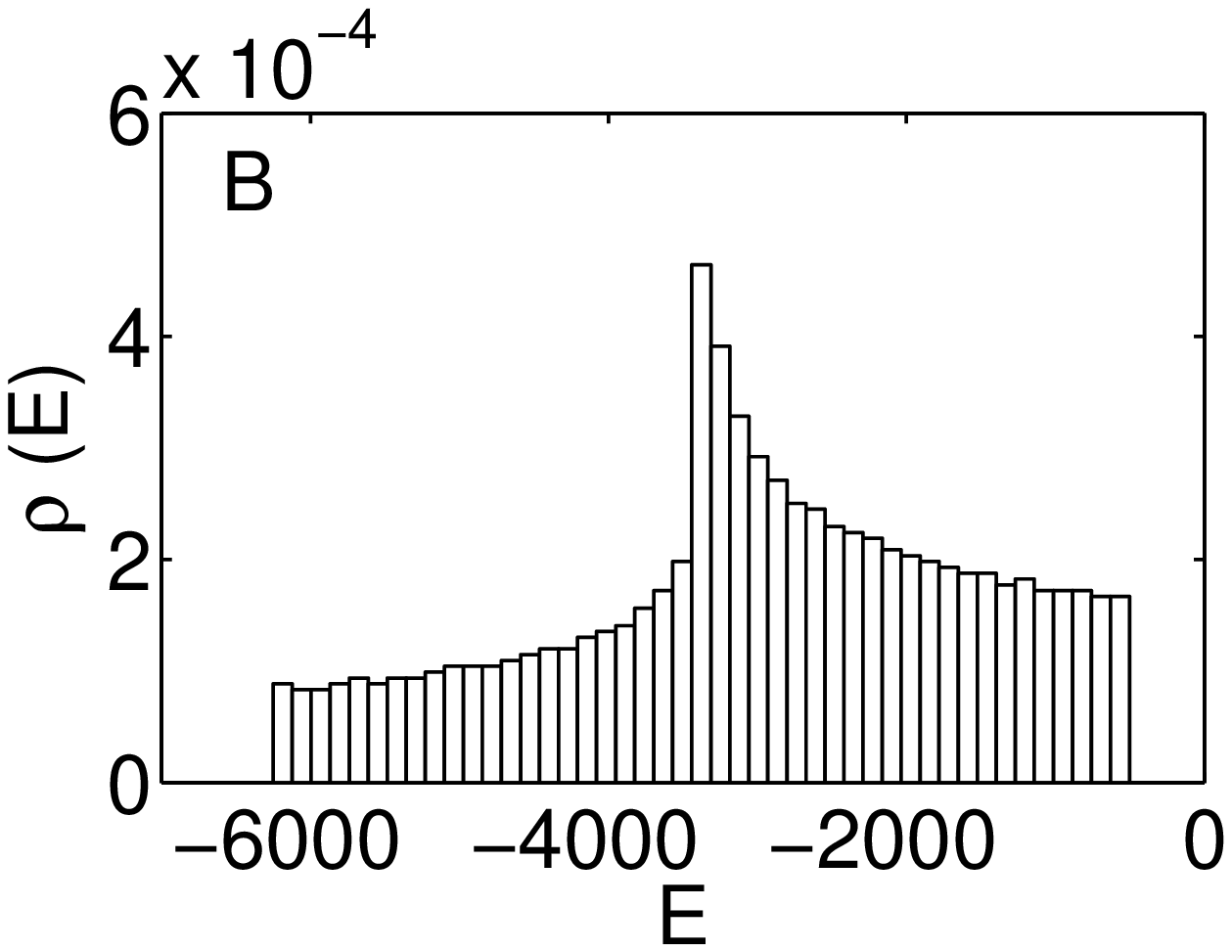}
\includegraphics[width=4.2cm]{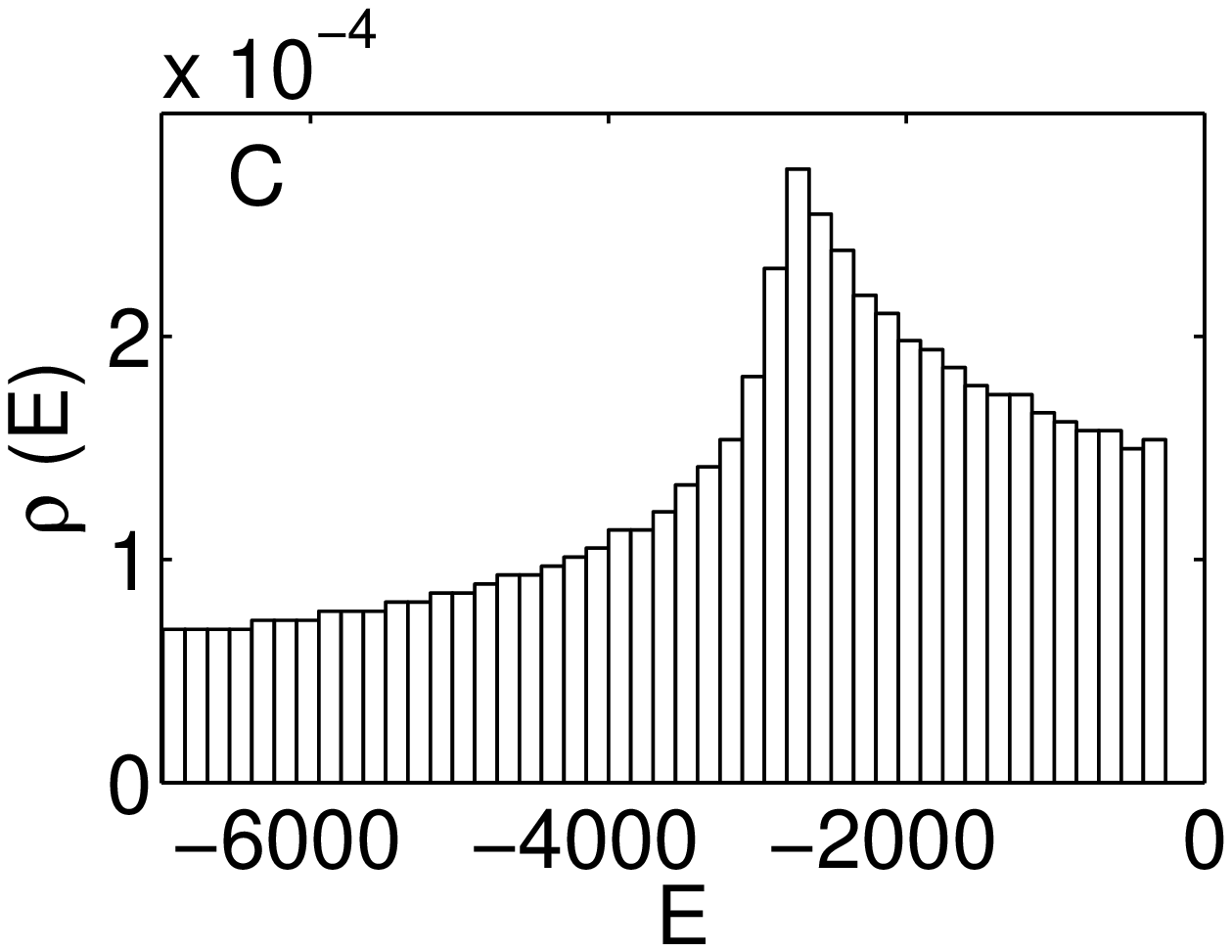}
\caption{\label{fig-density}
Level density $\varrho (E)$ of the many particle system in comparison to the mean-field
energies for $N=1500$ particles for $v=1$, $g=-3/N_s$ and $\varepsilon=0.5, 1, 1.5$ ($\hbar=1$).}
\end{figure}

\begin{figure}[b]
\centering
\includegraphics[width=6.6cm]{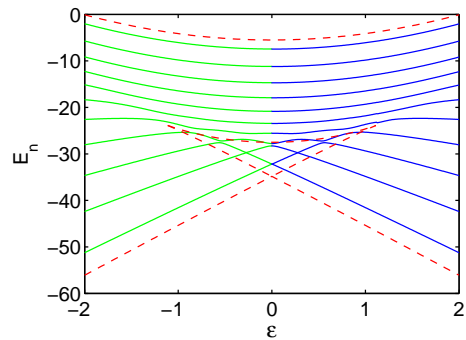}
\caption{\label{fig-levels_mp_mf-2}
(Color online) Many particle energies $E_n$ and mean-field eigenenergies ${\cal H}_{stat}$ (\textcolor{red}{- -})
as a function of the onsite energy $\varepsilon$
in the supercritical  region ($g=-3/N_s$) for $v=1$ and $N=10$ particles ($\hbar=1$).
The spectrum is organized by the mean-field eigenenergies ${\cal H}_{stat}$ (\textcolor{red}{- -}).
The exact quantum eigenvalues shown $\varepsilon\leq 0$ (\textcolor{green}{---}) are almost exactly reproduced by the
semiclassical ones, $E_n^{sc}$, shown for $\varepsilon\geq 0$ (\textcolor{blue}{---}).}
\end{figure}

Figure~\ref{fig-levels_mp_mf-1} shows an example of the many particle eigenvalues $E_n$ 
in dependence on $\varepsilon$ for a subcritical interaction strength. The pattern of eigenvalues varies
smoothly with $\varepsilon$ and is bounded by the stationary mean-field energies 
shown as red curves. Because of the symmetry of the spectrum for $\varepsilon
\rightarrow - \varepsilon$ the exact spectrum is only shown for $\varepsilon \le 0$, whereas
for  $\varepsilon \ge 0$ the semiclassical eigenvalues are shown as discussed below.
Figure \ref{fig-levels_mp_mf-2} shows a similar plot in the supercritical region. Here we
observe a net of avoided crossings
clearly organized by a skeleton provided by the stationary mean-field energies, as reported before 
by several authors \cite{Kark02,Buon04,06zener_bec}.  Again, for $\varepsilon \ge 0$ the
semiclassical eigenvalues are shown, which closely agree with the quantum ones in all
details. 

The mean-field eigenenergies (red curves) show a swallowtail structure which forms a caustic of the multi-particle
eigenvalue curves in the limit $N\rightarrow \infty$.
To illustrate this issue,
one can calculate the level density $\varrho (E)$ (normalized to unity)
as a function of the energy \cite{Aubr96}.
Figure~\ref{fig-density} shows a histogram of  the level density for $N=1500$ particles
and different values of $\varepsilon$.
The mean-field swallowtail curve between the cusps manifests itself
as a peak in the density of the many particle energies.
In the limit $N\to\infty$ the density $\varrho (E)$ approaches a smooth curve and the peak
develops into a singularity.
At the positions of the other mean-field eigenenergies one observes
finite steps. Indeed the quantum level densities shown in Fig.~\ref{fig-density} for
a large value of $N$ are directly related to the classical period $T$ of motion
by $T=\rd S/\rd E$, where $S$ is the classical action, which we will focus on in more detail in the following.
The height of the steps in the density plots
are simply given by the period of harmonic oscillation in the vicinity of the extrema and the singularity
corresponds to the separatrix motion.

\section{Semiclassical quantization}
\subsection{The classical action}
The most important ingredient of a semiclassical quantization is the
action $S(E)$, i.e.~the phase space
area enclosed by the directed curve ${\cal H}(p,q)=E$. The action $S(E)$ increases with $E$
from zero at the minimum energy of ${\cal H}(p,q)$ to $2\pi N_s\hbar$, the total available
phase space area, at the maximum energy of ${\cal H}(p,q)$.

\begin{figure}[htb]
\centering
\includegraphics[width=6.6cm]{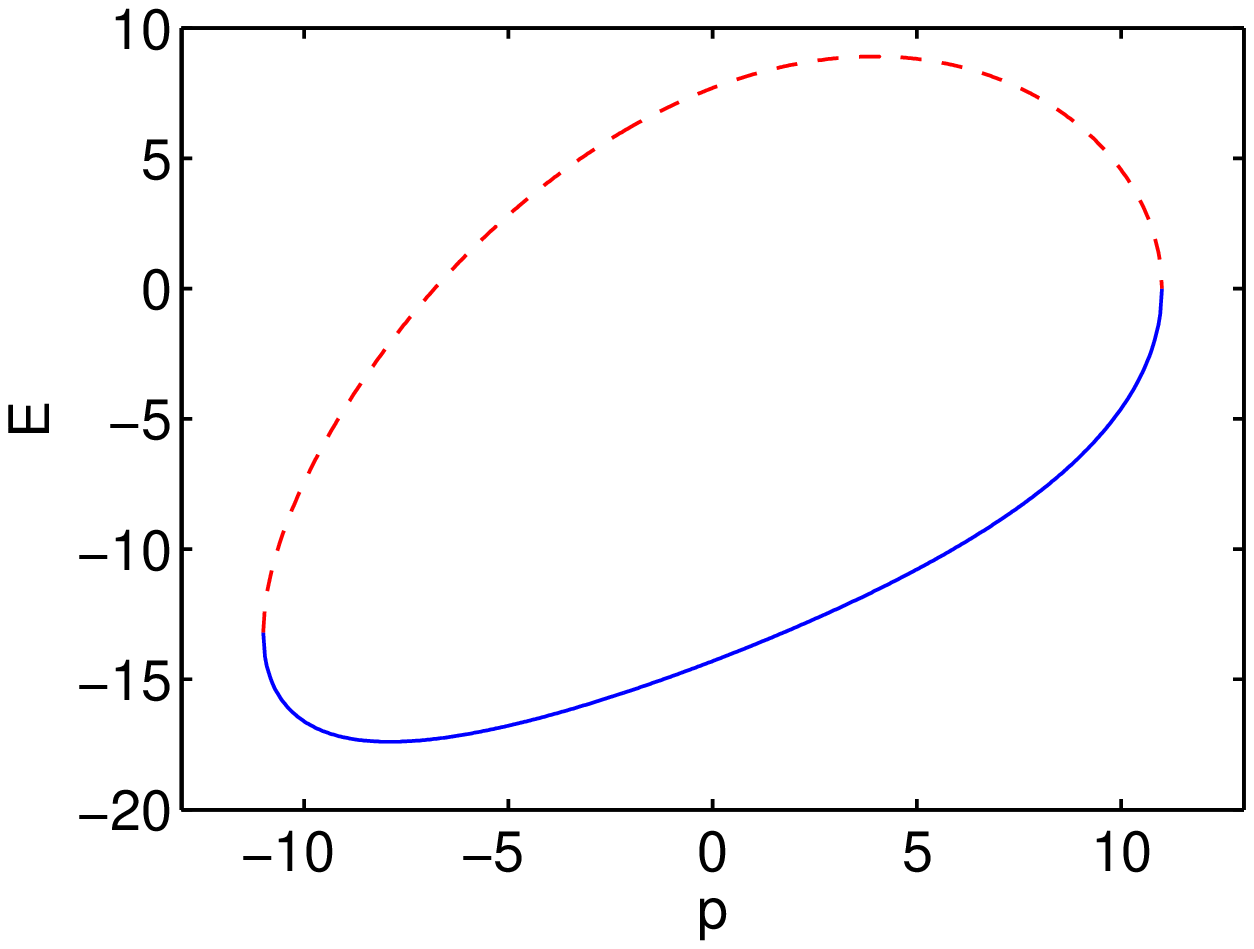}
\includegraphics[width=6.6cm]{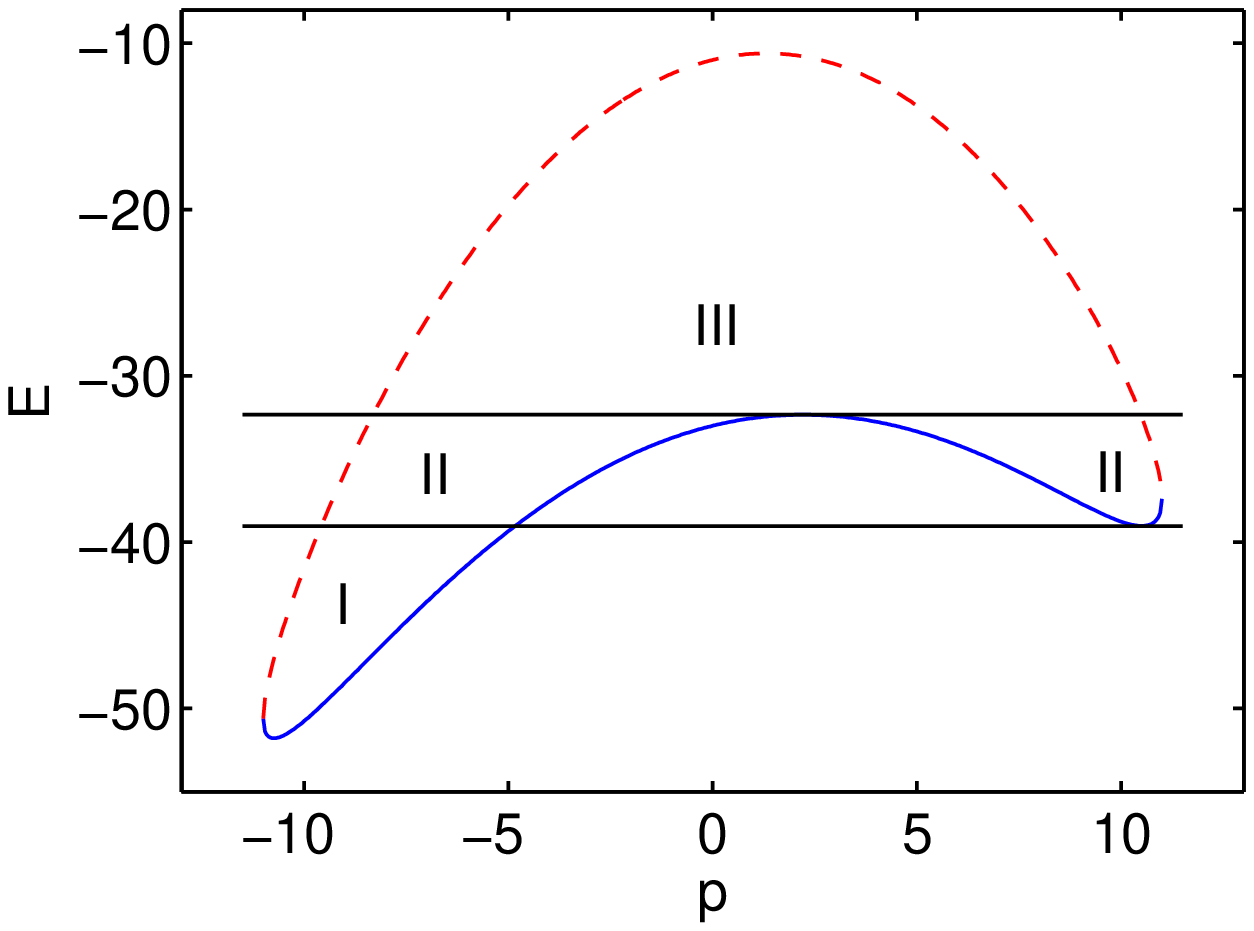}
\caption{\label{fig-Upm}
(Color online) The potentials $U_-(p)$ (\textcolor{blue}{---}) and $U_+(p)$ (\textcolor{red}{- -}), 
for $N=10$ particles, $v=1$ and $\varepsilon=0.6$
in the subcritical ($g=-0.6/N_s$, top) and
supercritical ($g=-4/N_s$, bottom) region, with $\hbar=1$.}
\end{figure}

For the generalized pendulum Hamiltonian (\ref{klHam}),
one can express the position variable $q$ uniquely as a function of $p$ and $E$ and write down the
action in momentum space in the form $S(E)=\oint q(p,E)\,\rd p$.
It is convenient \cite{Brau93,Haak01} to
introduce two functions $U_+(p)={\cal H}(p,0)$ and  $U_-(p)={\cal H}(p,\pi/2)$, which
join smoothly at $p=\pm \hbar N_s$ and
act as a potential for the variable $p$.
The classically allowed energy region is given by $U_-(p)\le E\le U_+(p)$
as illustrated in Fig.~\ref{fig-Upm} in the sub- and supercritical regions.
For a given energy $E$ the classical turning points
$p_\pm$ (with $p_- \le p_+$) are determined by $U_-(p_\pm)=E$
or $U_+(p_\pm)=E$.
One has to distinguish three basic types of motion and, with
$\widetilde S=\int_{p_-}^{p_+}q(p,E)\,\rd p$, we find:
\begin{enumerate}
\item Orbits encircling a minimum of ${\cal H}(p,q)$. The classical
turning points both lie on $U_-$ and we have $S(E)=\pi(p_{+}-p_{-})-2\widetilde S$.
\item Orbits encircling a maximum of ${\cal H}(p,q)$. The classical
turning points both lie on $U_+$ and we have
$S(E)=2\pi N_s\hbar-2\widetilde S$.
\item Rotor orbits extending over all angles $q$. We can find $p_-$ on $U_+$ and
$p_+$ on $U_-$ with $S(E)=\pi(N_s\hbar+p_{+})-2\widetilde S$
or $p_-$ on $U_-$ and
$p_+$ on $U_+$ with $S(E)=\pi(N_s\hbar-p_{-})-2\widetilde S$.
\end{enumerate}

\subsection{Energy quantization}
In the case of a single classically accessible region,
where there are two real turning points for any energy $E$,
the semiclassical quantization condition is given by
\be
S(E)=h (n+{\textstyle \frac12})\ ,\quad n=0,\,1,\,\ldots, N \,.\label{sem-quant1}
\ee
This simple case is always met in the subcritical region $|g|<v/N_s$.  
A numerical solution of (\ref{sem-quant1})
determines the semiclassical energies $E_n$, $n=0,\ldots, N$, where the
total available phase space area,
$0\le S(E)\le hN_s$, restricts the number of semiclassical
eigenvalues to $N_s$, exactly as the quantum ones. The resulting
semiclassical eigenvalues shown in Fig.~\ref{fig-levels_mp_mf-1}
for $N=10$ particles ($g=-0.5/N_s$, $v=1$, $\hbar=1$)
are in excellent agreement with the exact quantum ones. 

It should be pointed out, that in the noninteracting case, $g=0$,
the action $S(E)$ is a linear function of the energy $E$,
and the semiclassical eigenvalues agree with the exact ones
\be
E_n=\sqrt{\varepsilon^2+v^2}(2n-N)\ ,\quad n=0,\,1,\,\ldots, N \,.
\ee
This can be easily understood by recognizing that in this case
the Hamiltonian (\ref{BH-hamiltonian}) describes nothing but a system of
two coupled harmonic oscillators,
which can be transformed to two uncoupled ones by introducing normal coordinates.
It may also be of interest to note that (for $g=0$ and $N=2$ or $3$) the 
classical analog (\ref{klHam}) to the quantum Hamiltonian (\ref{BH-hamiltonian})
has been suggested many years ago by Miller and coworkers and
applied in a semiclassical description of electronic transitions
in atom-molecule collisions \cite{Mill78,Meye79}.

The supercritical region $|g|>v/N_s$ is more complicated. Here the energy surface 
has two minima,
hence the potential function $U_-(p)$ has two minima as well,
separated by a potential barrier.
In this case one has to distinguish different regions of the energy.
For energies below the upper minimum (region I in Fig.~\ref{fig-Upm}),
the quantization can be carried out like in the subcritical case by equation (\ref{sem-quant1}).
For energies between the upper minimum and the barrier $E_{\rm barr}$ (regions II in
Fig.~\ref{fig-Upm}),
there are four real turning points $p^{(l)}_{-}<p^{(l)}_{+}<p^{(r)}_{-}<p^{(r)}_{+}$.
In this case one has
to consider tunneling through the barrier.
The semiclassical quantization condition can be achieved
by a more elaborate expression \cite{Chil74} (see also \cite{Froe72,Chil91}):
\be
\sqrt{1+\kappa^2}\,\cos (S_l+S_{r}-S_\phi)=-\kappa \,\cos (S_r-S_{r}+S_\theta)\label{sem-quant2}
\ee
where $S_l$ and $S_{r}$ are the action integrals in the left resp. right region in Fig.~\ref{fig-Upm}
(note that also here one has to distinguish the different cases a) and c)).
The term
\be
\kappa =\re^{-\pi S_\epsilon}\ ,\
S_\epsilon=\frac{1}{\pi}\int_{p_+^{(l)}}^{p_-^{(r)}}|q(p,E)|\,\rd p
\ee
accounts for tunneling through the barrier,
\be
S_\phi=\arg \Gamma({\textstyle \frac12}+\ri S_\epsilon)-S_\epsilon\,\log |S_\epsilon| +S_\epsilon
\ee
is a phase correction, and $S_\theta=0$ below the barrier. Deep below the barrier, tunneling can be neglected and the simple semiclassical single well quantization is recovered (see also \cite{Wu06}).

Above the barrier, the inner turning points $p_+^{(l)}$, $p_-^{(r)}$ turn into
a complex conjugate pair and different continuations of the semiclassical quantization have been
suggested  \cite{Chil74,Froe72,Chil91}. Following \cite{Chil74} we replace these turning points
by the momentum at the barrier $p_{\rm barr}$ in the formulas for $S_{l,r}$, modify
the tunneling integral $S_{\epsilon}$ as
\be
S_\epsilon=\frac{\ri}{2}(p_-^{(r)}-p_+^{(l)})-\frac{\ri}{\pi}\int_{p_+^{(l)}}^{p_-^{(r)}}q(p,E)\,\rd p
\ee
and introduce a non-vanishing action integral
\be
S_{\theta}=\int_{p_+^{(l)}}^{p_{\rm barr}}q(p,E)\,\rd p  + \int_{p_-^{(r)}}^{p_{\rm barr}}q(p,E)\,\rd p.
\ee
The combined semiclassical approximation is continuous if the energy
varies across the barrier (from region II to III in Fig.~\ref{fig-Upm}) and continuously
approaches the simple version with only two turning points $p_-^{(l)}$ and $p_+^{(r)}$ in region III
high above the barrier.

\begin{figure}[htb]
\centering
\includegraphics[width=6.6cm]{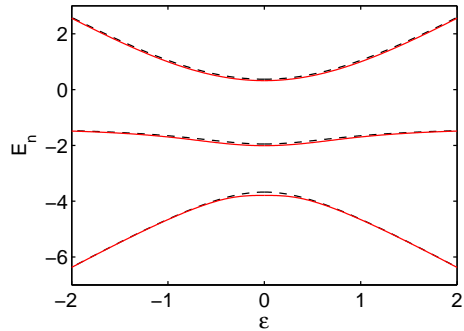}
\caption{\label{fig-levels_mp_mf-N2}
(Color online) Exact (- -), $E_n$ and semiclassical (\textcolor{red}{---}) many particle energies, $E_n^{sc}$, as a function of the onsite energy $\varepsilon$
in the subcritical  region ($g=-0.9/N_s$) for $v=1$ and $N=2$ particles ($\hbar=1$).}
\end{figure}

Figure \ref{fig-levels_mp_mf-2} shows the semiclassical  many particle energy eigenvalues
in dependence on the parameter $\varepsilon$
in the supercritical region for $N=10$ particles ($g=-3/N_s$, $v=1$, $\hbar=1$). 
Also here one observes an almost precise agreement with the exact eigenvalues and the net of
avoided level crossings in all details.
In particular the level distances at the avoided crossings are reproduced and allow furthermore
a direct semiclassical evaluation.
Figure \ref{fig-levels_mp_mf-N2} shows both exact and semiclassical 
eigenvalues in dependence on $\varepsilon$ for subcritical interaction 
for only $N=2$ particles. Even for that small particle number the semiclassical eigenvalues 
approximate the exact ones quite well. The deviation between the semiclassical 
and exact quantum eigenvalues decreases with increasing particle number $N$. 

For a more quantitative comparison, the exact and semiclassical eigenvalues are listed in 
Table~\ref{table1} for $N=20$ particles and selected $\varepsilon$-values. 
Here the relative error is only about $5\cdot10^{-4}$.

\begin{table}[t]
\begin{center}
\begin{tabular}{cc|cc|cc|cc}
\multicolumn{2}{c|}{$\varepsilon = 0$} &\multicolumn{2}{|c|}{ $\varepsilon = 0.5$} &
\multicolumn{2}{|c|}{$\varepsilon = 1.0$} &\multicolumn{2}{|c}{ $\varepsilon = 1.5$}\\ 
\rule[-2mm]{0mm}{6mm}  $E_n$ & $E_n^{sc}$ & $E_n$ & $E_n^{sc}$ &  $E_n$ & $E_n^{sc}$ & $E_n$ & $E_n^{sc}$   \\ \hline
   12.481 &  12.469 &  11.823 &  11.811 &   9.859 &   9.846 &   6.618 &   6.600 \\
   16.354 &  16.342 &  15.692 &  15.679 &  13.715 &  13.702 &  10.458 &  10.440\\
   20.097 &  20.085 &  19.429 &  19.417 &  17.437 &  17.424 &  14.161 &  14.143\\
   23.707 &  23.695 &  23.032 &  23.020 &  21.021 &  21.008 &  17.722 &  17.703\\
   27.178 &  27.167 &  26.496 &  26.484 &  24.462 &  24.449 &  21.135 &  21.115\\
   30.508 &  30.497 &  29.815 &  29.804 &  27.753 &  27.741 &  24.391 &  24.370\\
   33.690 &  33.679 &  32.985 &  32.974 &  30.888 &  30.875 &  27.481 &  27.458\\
   36.718 &  36.708 &  35.997 &  35.987 &  33.857 &  33.844 &  30.395 &  30.369 \\
   39.585 &  39.575 &  38.845 &  38.835 &  36.648 &  36.635 &  33.115 &  33.085\\
   42.281 &  42.272 &  41.516 &  41.507 &  39.246 &  39.234 &  35.622 &  35.583\\
   44.795 &  44.786 &  43.999 &  43.990 &  41.630 &  41.618 &  37.896 &  37.829\\
   47.111 &  47.104 &  46.273 &  46.265 &  43.758 &  43.745 &  40.090 &  40.070\\
   49.181 &  49.176 &  48.301 &  48.299 &  45.649 &  45.642 &  43.015 &  43.023\\
   51.112 &  51.107 &  50.031 &  50.024 &  46.729 &  46.739 &  46.847 &  46.853\\
   52.193 &  52.192 &  51.406 &  51.406 &  48.952 &  48.979 &  51.439 &  51.443 \\
   54.690 &  54.687 &  52.871 &  52.870 &  52.760 &  52.771 &  56.717 &  56.720\\
   54.783 &  54.781 &  54.680 &  54.678 &  57.413 &  57.419 &  62.641 &  62.643\\
   58.828 &  58.825 &  56.738 &  56.750 &  62.782 &  62.786 &  69.187 &  69.188\\
   58.829 &  58.826 &  61.512 &  61.518 &  68.813 &  68.815 &  76.340 &  76.341\\
   63.766 &  63.763 &  67.009 &  67.013 &  75.475 &  75.477 &  84.090 &  84.091\\
   63.766 &  63.763 &  73.171 &  73.173 &  82.751 &  82.752 &  92.432 &  92.433\\
\end{tabular}
\caption{\label{table1} Exact quantum $E_{n=0...20}$ and semiclassical eigenvalues $E_n^{sc}$ for 
$\varepsilon = 0,\,0.5,\,1.0,\,1.5$ for  $v=1$ and $N=20$ particles ($\hbar=1$)
in the supercritical region ($g=-3/N_s$).}
\end{center}
\end{table}

\subsection{Eigenfunctions}
The mean-field approximation allows also a semiclassical construction of the 
eigenstates $\hat{H}|\Psi_n\rangle=E_n|\Psi_n\rangle$ of the Bose-Hubbard Hamiltonian (\ref{BH-hamiltonian}) 
resp.~(\ref{BH-hamiltonian-SR}). In the quantum case, the main interest may
concentrate on the population imbalance of these states, i.e. the $p$-representation
\be
|\Psi_n\rangle=\sum_{p=-N:2:N}\Psi_n(p)\,|p\rangle\,,
\ee
where $p$ runs from $-N$ to $N$ in steps of 2.
Based on the (classical) mean-field dynamics, we have to construct a 
semiclassical approximation in momentum space, which is discussed in some detail in 
\cite{00mom}.
The purely classical momentum probability distribution is easily calculated as
$w^{cl}(p)=(2T|\partial{\cal H}/\partial q|)^{-1}$, where $T$ is the period of oscillation.
Note that the factor 2 arises from the two classical contributions, i.e.~the direct
path and the path once reflected at the opposite turning point. For our
mean-field Hamiltonian (\ref{klHam}) this leads to
\be
\textstyle
w^{cl}(p)=C\Big[v^2\big(N_s^2-\frac{p^2}{\hbar^2}\big)-
\big(E^{sc}_n-\varepsilon \frac{p}{\hbar}-\frac{g}{2}(N_s^2
+\frac{p^2}{\hbar^2})\big)^2\Big]^{-\frac{1}{2}}
\ee
in the classically allowed region,
where $C=1/2T$ takes care of the normalization. The so-called 
primitive semiclassical wavefunction includes interference of the two
classical paths:
\be
 |\Psi_n^{sc}(p)|^2=2\,|w_c(p)|\,\cos ^2\big(S(p)/\hbar- \pi/4\big)
\ee
where $S(p)=S(p,E^{sc}_n)$ is the classical action for an energy equal to the
semiclassical eigenenergy $E^{sc}_n$ of state number $n$, i.e.~the 
oriented momentum-integral over $q(p)=q(p,E^{sc}_n)$
\be\label{eqn-action-wave}
 S(p)=-\int_{p_-}^p{q(p')\,dp'}+\frac{\pi}{2}\big(p-p_-\big),
\ee
if $p_-$ lies on the lower potential curve $U_-$ or
\be
 S(p)=\int_{p_-}^p{q(p')\,dp'}.
\ee
if $p_-$ lies on $U_+$. In the classical forbidden region $q(p)$ is
complex valued and we can use the usual complex continuation \cite{Chil91}
\be
|\Psi_n^{sc}(p)|^2=\frac{1}{2}\left|w^{cl}_n(p)\exp{(-2\ri S(p)/\hbar)}\right|;
\ee
where 
\be
S(p)=\left\lbrace\begin{matrix}
\mp \int_p^{p_{-}}{q(p)\,dp},\qquad p<p_-,\ p_-\ \text{on}\ U_\pm\\[2mm]
\mp \int_{p_+}^{p}{q(p)\,dp},\qquad p>p_+,\ p_+\ \text{on}\ U_\pm
                                     \end{matrix}\right.\,.
\ee 
Note that these distributions should be renormalized to unity. 

\begin{figure}[htb]
\centering
\includegraphics[width=6.6cm]{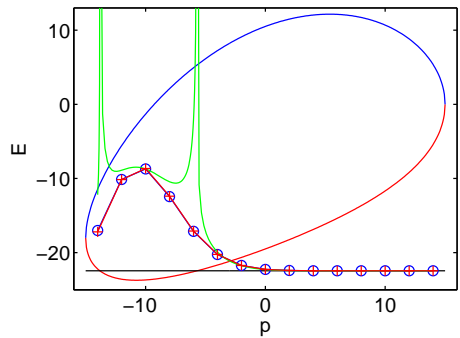}
\caption{\label{fig-wavefun0}
(Color online) Momentum ``potentials'' $U_\pm(p)$ and -- plotted
at the energy $E_n$ -- 
exact (blue circles), primitive semiclassical (green) and 
uniform semiclassical (red crosses) wavefunctions $|\Psi_n(p)|^2$
of the lowest eigenstate $n=0$ 
for $N=14$ particles ($g=-0.6/N_s$, $\varepsilon=0.6$, $v=1$ and $\hbar=1$). 
The solid lines are drawn to guide the eye.}
\end{figure}

\begin{figure}[htb]
\centering
\includegraphics[width=6.6cm]{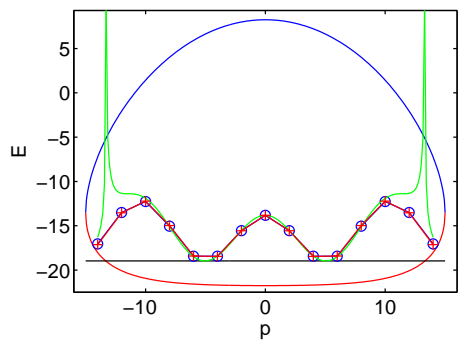}
\caption{\label{fig-wavefun2}
(Color online) Momentum ``potentials'' $U_\pm(p)$ and -- plotted
at the energy $E_n$ -- exact (blue circles), primitive semiclassical (green) and uniform
semiclassical (red crosses) wavefunctions $|\Psi_n(p)|^2$ of eigenstate $n=2$ for 
$N=14$ particles
($g=-0.9/N_s$, $\varepsilon=0$, $v=1$ and $\hbar=1$). The solid lines are drawn 
to guide the eye.}
\end{figure}

The divergence at the classical turning points $p_\pm$ is finally cured
by a uniform semiclassical approximation (see e.g.~\cite{Chil91}). 
Here the different turning point
scenarios discussed above must be distinguished.In the following we only state the
result if $p_\pm$ both lie on the lower potential curve $U_-(p)$. A mapping
onto harmonic oscillator wave functions then yields \cite{Chil91}
\be
|\Psi_n^{sc}(p)|^2\sim \left|w^{cl}_n(p)\sqrt{2n+1-\xi^2}\right|\text{H}_n(\xi)^2\exp{(-\xi^2)}
\label{uniform}
\ee
where H$_n$ denotes the Hermite polynomial of $n$th order 
and $\xi$ is determined by
\be
\frac{1}{2}\xi\,\sqrt{\xi_0^2-\xi^2}+\frac{1}{2}\xi_0^2\Big(\frac{\pi}{2}+
\arcsin{\big(\frac{\xi}{\xi_0}\big)}\Big)=S(p)
\ee
with $\xi_0=\sqrt{2n+1}$. 

Up to now, the semiclassical momentum variable $p$ could be treated as continuous
in the mean-field approximation, whereas in the quantum system, $p$ is a discrete
variable, $p=-N,-N+2,\ldots,+N$. Semiclassically, this is a consequence of the
even symmetry and the $\pi$-periodicity of the mean-field Hamiltonian 
(\ref{klHam}) in the coordinate $q$. As in Fourier transformation, this allows only 
even or odd integer values of $p$. 

The final uniform semiclassical wave functions in momentum space are therefore
given by (\ref{uniform}) at $p=-N,\ldots,N$, 
normalized as $\sum_{p=-N:2:N}|\Psi_n^{sc}(p)|^2=1$.

Figures \ref{fig-wavefun0} and \ref{fig-wavefun2} show a comparison of
the primitive semiclassical approximation (normalized to fit the central 
maximum) and the uniform one with
exact quantum results, both in the subcritical region for $N=14$ particles.
Shown is the ground state $n=0$ for a biased Bose-Hubbard model 
($\varepsilon=0.6$) and the third excited state $n=2$ for a symmetric
one ($\varepsilon=0$). As expected, the quantum distributions 
mainly populate the classically allowed region inside the ``potential'' curves
$U_\pm(p)$ and are very well approximated by the primitive semiclassical
distributions. In particular, the uniform approximation is almost indistinguishable
from the exact values.

\section{Conclusion}
It is demonstrated for a two-mode Bose-Hubbard model, that the
mean-field approximation can be used to reconstruct approximately the individual eigenvalues 
in a semiclassical Bohr-Sommerfeld (or EBK) manner with astounding 
accuracy even for a small number of particles. 
The same holds for the primitive semiclassical approximation of corresponding eigenstates 
which was shown for the subcritical case. Furthermore the possibility of a uniform approximation was 
demonstrated for a special case.

For the two-mode Bose-Hubbard system considered here, the classical
description provided by the mean-field
model has one degree of freedom and is therefore
integrable. For
three and more modes, the classical dynamics is chaotic (see, e.g., the
studies of the three-mode system \cite{Moss06,05level3}
or tilted optical lattices \cite{Thom03}). Chaoticity also appears in
periodically driven two-mode systems \cite{Holt01a,07kicked} or the related kicked tops
\cite{Haak01}.
A semiclassical description of the quasienergy spectrum in these cases is
a challenge for future studies.

Finally it should be noted that the semiclassical analysis used in the
present paper is based on well-known results which allow, e.g., a
straightforward treatment of tunneling corrections. Basically these
theories are, however, valid for a flat phase space. More recent
developments directly address semiclassical quantization of spin
Hamiltonians with a compact phase space (see, e.g., \cite{Shan80,Garg04,Nova05}  
and references given there). This research is, however, still in progress
and applications to Hamiltonians like (\ref{BH-hamiltonian-SR}) 
including tunneling corrections will be the topic of future investigations.

\begin{acknowledgments}
Support from the Deutsche
Forschungsgemeinschaft via the Graduiertenkolleg ''Nichtlineare Optik
und Ultrakurzzeitphysik'' is gratefully acknowledged.
\end{acknowledgments}


\begin{thebibliography}{10}
\bibitem{semiMP_Bem1}
Note that the energy spectrum stays the same if the sign of $v$ is altered. If
  the sign of $g$ is altered it is turned upside down. This does not change the
  subsequent approach in principle.
\end{thebibliography}
\end{document}